\documentclass[11pt]{article}

\usepackage{amsthm}
\usepackage{a4wide}
\usepackage{longtable}

\def\it{\textit} 
\def\bf{\textbf}

\def\mc{\mathcal}
\def\mb{\mathbf}

\newcommand{\be}{\begin{equation}}
\newcommand{\ee}{\end{equation}}
\newcommand{\benum}{\begin{enumerate}}
\newcommand{\eenum}{\end{enumerate}}
\newcommand{\bit}{\begin{itemize}}
\newcommand{\eit}{\end{itemize}}

\newtheorem{thom}{Theorem}

\newtheorem{prop}[thom]{Proposition}

\newtheorem{defn}{Definition}

\begin{document}

\title{Syntactic Characterisations of Polynomial-Time Optimisation Classes
(Syntactic Characterizations of Polynomial-Time Optimization Classes)}

\author{P. Manyem\footnote{Centre for Informatics and Applied Optimisation, 
School of IT and Mathematical Sciences, 
University of Ballarat, 
Mount Helen, VIC 3350, Australia. 
mailto: p.manyem@ballarat.edu.au}}

\date{}

\maketitle{\thispagestyle{empty}}

\begin{abstract}
In Descriptive Complexity, there is a vast amount of literature on
decision problems, and their classes such as \textbf{P, NP, L and NL}. ~
However, research on the descriptive complexity of optimisation problems
has been limited.  Optimisation problems corresponding to the
\textbf{NP} class have been characterised in terms of logic expressions
by Papadimitriou and Yannakakis, Panconesi and Ranjan, Kolaitis and
Thakur, Khanna et al, and by Zimand.  Gr\"{a}del characterised the
polynomial class \textbf{P} of decision problems.  In this paper, we
attempt to characterise the optimisation versions of \textbf{P} via
expressions in second order logic, many of them using universal Horn
formulae with successor relations.  The polynomially bound versions of
maximisation (maximization) and minimisation (minimization) problems are
treated first, and then the maximisation problems in the ``not
necessarily polynomially bound" class.
\end{abstract}

\setcounter{page}{1} 

\section{Introduction}\label{sec:intro}

Though there has been abundant research in Descriptive Complexity
since Fagin's 1974 theorem \cite{fagin} (which captures the class
NP as the set of properties that can be represented in existential
second order logic), the application of this area to approximation
complexity has been limited. Approximation complexity measures how
well an NP-hard optimisation problem can be approximated, or how
far is the value of a (possible) heuristic solution from that of
an optimal solution.

A few attempts to characterise approximation classes in terms of
logic are: Papadimitriou and Yannakakis in 1991
\cite{papaYanna91}, Panconesi and Ranjan in 1993
\cite{pancoRanjan}, Kolaitis and Thakur in 1994 and 1995 \cite{KT94,KT95}, and
Khanna et al in 1998 \cite{Kmsv98}.

The approximation complexity of a problem $P$ is usually measured
by the \it{approximation ratio} that a heuristic $H$ for $P$ can
guarantee, over all instances of $P$. ~The approximation ratio
$R_H(I)$ obtained by $H$ for a given instance $I$ of $P$ is given
by
\begin{equation}
R_H(I) = \frac{\mbox{value obtained by $H$ on $I$}}{\mbox{value of
an optimal solution for $I$}}
\end{equation}

In \cite{KT94,pancoRanjan,papaYanna91}, the authors characterise
approximation hardness in terms of quantifier complexity --- the
number and types of quantifiers that appear at the beginning of a
second-order formula in prenex normal form (PNF).  ~For a formula
in PNF, all quantifiers appear at the beginning, followed by a
quantifier-free formula.

\subsection{Contributions in this paper}
In this paper, we first present a logical representation of a
subclass of $\mb{P^{\prime}}$ --- $\mb{P^{\prime}}$ is the class of optimisation
problems that can be solved to optimality within polynomial
time\footnote{Strictly speaking, in Turing machine terminology,
$\mb{P^{\prime}}$ is the set of languages where, if an instance
$I$ of an optimisation problem $P \in \mb{P^{\prime}}$ is encoded
as an input string $x$ in some alphabet $\Sigma$, a deterministic
Turing machine will compute the optimal solution (which is again a
string) within $\Theta(|x|^k)$ steps, where $k$ is some constant
and $|x|$ is the length of the input string.}. The class of \it{decision
problems} corresponding to $\mb{P^{\prime}}$ is
\bf{P}. 
~The particular subclass $\mb{Q^{\prime}}$ (of $\mb{P^{\prime}}$) that we
focus on includes only \it{polynomially bound optimisation problems}, defined
below in Definition \ref{polyMax}.  In particular, we
\bit
\item
provide syntactic characterisations for both maximisation and
minimisation problems in $\mb{Q^{\prime}}$,
\item
give examples of characterisations (MAXFLOW$_{PB}$ for maximisation
and SHORTEST PATH$_{PB}$ minimisation),
\item
show that MAXFLOW$_{PB}$ is complete for the maximisation subclass of
$\mb{Q^{\prime}}$,
\item
present characterisations for maximisation problems in $\mb{P^{\prime}}$
(defined in Table \ref{tab:defines} --- problems not necessarily
polynomially bound), as well as an example for a problem in this class
(MAXIMUM MATCHING). ~This is the most significant contribution in this
paper, and is a considerable departure from the treatment in
\cite{zimand98}.  Whereas that paper studied problems in \bf{N} in
general, we study maximisation problems in $\mb{P^{\prime}}$. 
\eit

~The syntactic characterisation of \bf{P} is given below in
Theorem \ref{thom:decision}, as shown in Gr\"{a}del \cite{gradel91}.

\subsection{Notation and Definitions}

\begin{table}
\begin{center}
\begin{tabular}{|l|p{300pt}|}
\hline
$\sigma  $  & vocabulary  \\
\hline
$\bf{A}  $  & a
structure defined over $\sigma$
               (captures an instance of an optimisation problem) \\
\hline
 $\eta  $    & a quantifier-free first order formula, and a
conjunction of \it{Horn} clauses at the same time.
(Recall that a Horn clause contains at most one positive literal.)  \\
\hline
$\mb{x}  $  &  an $m-$tuple of first order variables \\
\hline
$\mb{S}  $  &  a sequence of second-order variables (predicate symbols) \\
            &  (captures a solution to the optimisation problem)
            \\
\hline
 $\mb{P}  $   &  computational class of \it{decision}
problems,
decidable in polynomial time by a deterministic Turing machine \\
\hline
 $\mb{P^{\prime}}$    &  class of \it{optimisation} problems
corresponding to $\mb{P}$   \\
                      &  (also called $\mb{P}-$optimisation problems) \\
\hline
$\mb{Q^{\prime}}$    &   $\mb{Q^{\prime}} \subseteq \mb{P^{\prime}}$, and
                       $\mb{Q^{\prime}}$ only contains \it{polynomially
                       bound}
                      optimisation problems   (see Definition \ref{polyMax})    \\
\hline
\bf{N}    &  class of \it{optimisation} problems whose decision
              versions are in \bf{NP}  \\
\hline
$\mb{N^{\prime}}$    &   $\mb{N^{\prime}} \subseteq \mb{N}$, and
                       $\mb{N^{\prime}}$ only contains \it{polynomially
                       bound} optimisation problems  \\
\hline
ESO      &  Existential Second Order Logic  \\
\hline
PNF      &  Prenex Normal Form              \\
\hline
\end{tabular}
\caption{Notation} 
\label{tab:defines}
\end{center}
\end{table}

All notation is defined in Table \ref{tab:defines}, as a {one-stop} 
reference point.  For the same reason, all definitions are provided
below in this section. 

\begin{defn}
An optimisation problem $Q^{\prime}$ is said to
be  \bf{polynomially bound} if the value of an optimal solution
to every instance $I$ of $Q^{\prime}$ is bound by a polynomial
in the size of $I$. ~In other words, there exists a
polynomial $p$ such that
\begin{equation}
opt_{Q^{\prime}} (I) \leq p(\vert I \vert),
\label{polyBoundDef}
\end{equation}
for {every} instance $I$ of $Q^{\prime}$.  The class of all such
problems is $\mb{Q^{\prime}}$. 
\label{polyMax}
\end{defn}

\begin{defn}
\bf{First order logic} consists of a
\it{vocabulary} (alias signature) $\sigma$, and \it{models} (alias
\it{structures}) defined on the vocabulary.  In its simplest form,
a vocabulary consists of a set of variables, and a set of relation
symbols $R_j (1 \leq j \leq J)$, each of arity $r_j$.  A model $M$
consists of a universe $U$ whose elements are the values that
variables can take --- $M$ also instantiates each relation symbol $R_j
\in \sigma$ with tuples from $U^{(r_j)}$.
For example, a model $\mb{G}$ in graph theory
may have the set of vertices $G = \{1, 2, \cdots 10\}$ as its
universe (assuming that the graph has 10 vertices), and a single
binary relation $E$ where $E(i,j)$ is true iff $(i,j)$ is an edge
in the graph $\mb{G}$.  ~A model represents an instance of an
optimisation problem.  
\label{FOL}
\end{defn}

\begin{defn}
A \bf{$\mb{\Pi_1}$ \textnormal{(}$\mb{\Sigma_1}$\textnormal{)}
first order formula in PNF} only has
universal (existential) quantifiers, quantified over first order variables.
\label{pie1}
\end{defn}

\begin{defn}
An \bf{existential second-order (ESO) Horn}
expression is of the form $\exists \mb{S} \psi$, where $\psi$ is a
first order formula, and $\mb{S} = (S_1, ~ \cdots ~ S_p)$ is a sequence
of predicate symbols not in the vocabulary of $\psi$.  The formula
$\psi$ can be written in $\Pi_1$ form as
\begin{equation}
\psi = \forall x_1 \forall x_2 \cdots \forall x_k \eta = \forall
\mb{x} ~ \eta.
\label{hornDef}
\end{equation}
where $\eta$ is a conjunction of Horn clauses ($\eta$ is, of
course, quantifier-free), and $x_i$ $(1 \leq i \leq k)$ are first
order variables.  Each clause in $\eta$ contains at most one positive
occurrence of any of the second order predicates $S_i$ ($1 \leq i \leq
p$). 
\label{ESOhorn}
\end{defn}

\begin{defn}
A \bf{$\mb{\Pi_2}$ \textnormal{(}$\mb{\Sigma_2}$\textnormal{)}
 formula in prenex normal form (PNF)} can be
written as follows: 
\begin{equation}
\phi = 
\forall x_1 \cdots \forall x_a ~
\exists y_1 \cdots \exists y_b ~
\eta ~~~~ 
(\phi = 
\exists y_1 \cdots \exists y_b ~
\forall x_1 \cdots \forall x_a ~
\eta )
\label{Pi2Def}, 
\end{equation}
where $\eta$ is quantifier-free, ~$a, b \geq 1$, ~and the 
$x$'s and $y$'s are first-order variables. 
\label{pie2sigma2}
\end{defn}
The following theorem is due to Gr\"{a}del \cite{gradel91} --- 
this is the polynomial-time counterpart of Fagin's
theorem \cite{fagin} which characterised the class \bf{NP}: 
\begin{thom}
For any ESO Horn expression as defined in Definition \ref{ESOhorn}, the
corresponding decision problem is in $\mb{P}$.

The converse is also true --- if a problem $P$ is in $\mb{P}$, then it
can be expressed in ESO Horn form  --- but only if a
successor relation is allowed to be included in the vocabulary of
the first-order formula $\psi$. 
\label{thom:decision}
\end{thom}

\section{Polynomially Bound Optimisation Problems $\mb{Q^{\prime}}$}\label{sec:polyBound}

\bf{Optimisation problems corresponding to P}. 
We assume that for a maximisation (or a minimisation) problem
$Q^{\prime}$ in the class $\mb{Q^{\prime}}$ (corresponding to the class
\bf{P} of decision problems), the following can be computed in
polynomial time deterministically: 
(a) The value of the objective function $f(\mb{A},\mb{S})$ to a
solution \bf{S} of an instance \bf{A}, and 
(b) Whether a solution \bf{S} is a feasible solution to an instance
\bf{A}. 

We will study maximisation problems first, and then the minimisation
problems. 

\subsection{Polynomially Bound P-Maximisation Problems}
\label{sec:polyBoundMax}

For maximisation problems in $\mb{N^{\prime}}$ (see Table
\ref{tab:defines}
for a definition of $\mb{N^{\prime}}$), Kolaitis and Thakur
\cite{KT94} proved the following:
\begin{thom}
A maximisation problem $Q \in \mb{N^{\prime}}$ if and only if there
exists a $\Pi_2$ first order formula $\phi (\mb{w}, \mb{S})$  with
predicate symbols from the vocabulary $\sigma$ (of $\phi$) and the
sequence $\mb{S}$, such that for every instance $\mb{A}$ of $Q$,
the optimal solution value is given by 
\begin{equation}
opt_{Q} (\mb{A}) = \max_{\mb{S}} \vert \left\{ \mb{w}:
(\mb{A}, \mb{S}) \models \phi (\mb{w},\mb{S}) \right\} \vert. 
\end{equation}
\label{thom:KT94max}
\end{thom}

In other words, polynomially bound NP-maximisation problems fall in
what is called the MAX $\Pi_2$ class. 
We can show a similar result for 
the polynomial-time counterpart of $\mb{N^{\prime}}$, that is, 
maximisation problems in $\mb{Q^{\prime}}$:

\begin{thom}
 Let $\mb{A}$ be a structure
(instance) defined over $\sigma$.  The value of an optimal
solution to an instance $\mb{A}$ of a maximisation problem
$Q^{\prime}$ can be represented by
\begin{equation}
opt_{Q^{\prime}} (\mb{A}) = \max_\mb{S} \vert \{\mb{w}: (\mb{A},
\mb{S}) \models \forall \mb{x} ~ \eta(\mb{w}, \mb{x}, \mb{S}) \}
\vert 
\label{maxDef}
\end{equation}
if $Q^{\prime} \in \mb{Q^{\prime}}$,
where $\mb{x}$, $\mb{A}$, $\mb{S}$ and $\eta$ are defined in
Table \ref{tab:defines}.
\label{thom:hornMax}
\end{thom}

\begin{proof}

Let $Q$ and ${Q^{\prime}}$ be the decision and optimisation versions
respectively. 

We first show that $Q^{\prime}$ is polynomially
bound.  For this, the number of tuples $\mb{w}$ in an optimal
solution $\mc{S^{\ast}}$ should be polynomial in $\vert {A}
\vert$, the size of the universe of $\mb{A}$. ~It suffices to show
this for any solution $\mc{S}$.  ~Recall from Table
\ref{tab:defines} that the sequence \bf{S} of predicates captures a
(corresponding) solution $\mc{S}$ to the optimisation problem.

Suppose $\mb{w}$ is a $\mc{R}-$dimensional tuple. For a given universe
$A$ of $\mb{A}$, the number of possible tuples is $|A|^\mc{R}$
--- this is true for any solution $\mc{S}$ including the optimal one.
~Hence $Q^{\prime}$ is polynomially bound.

To complete the proof, we should show that if 
$Q^{\prime} \in \mb{Q^{\prime}}$, then the optimal solution value to
an instance $\mb{A}$ of $Q^{\prime}$ can be represented by equation
(\ref{maxDef}).  

Refer to Gr\"{a}del's theorem (Theorem
\ref{thom:decision}).  The decision problem $Q$ can be written as
an ESO Horn expression $\exists \mb{S} \psi$, except that now, $\psi$
should include a successor relation in its vocabulary, in addition
to being a Horn first order formula.   Problem $Q$ can be posed
as: ~Given an instance (a finite structure) $\mb{A}$, is there a
feasible solution $\mb{S}$ such that $f(\mb{A}, \mb{S}) \geq K$,
where $K$ is a certain integer ? \\
 (Here $f$ is the value of the objective function to solution $\mb{S}$
for the optimisation problem. Assume that we deal only with
problems with integer-valued objective functions.) 

A feasible solution $\mb{S}$ could consist of several relations
$S_1, S_2, \cdots S_p$ of arities $r_1, r_2, \cdots r_p$.  The
formula $\psi$ should be able to express $f(\mb{A}, \mb{S}) \geq
K$ --- however, this is insufficient.  Given a solution $\mb{S}$, we
know that all feasibility conditions, including $f(\mb{A}, \mb{S}) \geq
K$, can be checked in polynomial time deterministically for all problems 
in \bf{NP}. 
~What distinguishes \bf{P} from \bf{NP} is the fact that for
optimisation problems corresponding to \bf{P}, the optimal solution
value 
$\displaystyle opt_{Q^{\prime}} (\mb{A}) $
 can be computed in polynomial time deterministically, whereas
for optimisation problems corresponding to \bf{NP}, we only know that
this value can be computed in polynomial time non-deterministically. 

Hence this condition should be modified to 
$\displaystyle g(\mb{A}) = opt_{\mb{S}} f(\mb{A}, \mb{S}) \geq K$, where 
$g(\mb{A})$ is the optimal solution value to
instance \bf{A} over all solutions \bf{S}.
~Each of the $g(\mb{A})$ number of entities can be considered to be a
tuple $\mb{w_i}$, and thus we need at least $K$ such tuples.  These
tuples, at least $K$ in number, 
can be defined to form a new relation $F$ (on the universe $A$ of
$\mb{A}$) of arity
$k$.  Thus we want $|F|$, the number of tuples \bf{w} that satisfy
$F(\mb{w})$, to be at least $K$. 

\bf{Digression to discuss arity k}. As examples, setting $k = 2$ will suffice 
for the LONGEST PATH problem where the number of
arcs in a path is to be maximised, and $k = 1$ in a MAXSAT problem where the
number of satisfying clauses is to be maximised.  However, this can
handle only up to very small values of the objective function.  In the
LONGEST PATH case, we can only count $|A|^2$ tuples at most (where $|A|$ is
the number of vertices in the graph). However, if arc lengths are higher
than one, but still polynomially bound in $|A|$, the length of the longest path
--- though still polynomially bound in $|A|$ --- 
could be well above $|A|^2$, and this length cannot be handled by an arity
of $k = 2$ --- a higher arity is required.  Hence it would be safest to
increase the arity to $\mc{R}$, since $|A|^\mc{R}$ is the upper bound on
the objective function value.  A similar argument applies to the
weighted MAXSAT problem with polynomially bound weights --- and to all
polynomially bound NP-maximisation problems in general. 

Recall from Theorem \ref{thom:decision} that $\psi$ is in $\Pi_1$
form, where the quantifier-free part of $\psi$ is a conjunction of Horn
clauses, each of which contains at most one positive occurrence of
\it{any}\footnote{
For example, if $S_1$ and $S_2$ are second order predicates, then a Horn
clause cannot contain both $S_1$ and $S_2$ as positive literals.} 
of the relation symbols $S_i$. 
~Hence $\psi$ can be written as 
$\forall x_1, \cdots \forall x_m \hat{\eta}$, where $\hat{\eta}$ is an
expression consisting of variables $x_1, \cdots x_m$, all predicates
from $\mb{S}$, and the the relation $F$. ~That is,
\begin{equation}
\psi = \forall x_1 \cdots \forall x_m 
~ \hat{\eta} (x_1, \cdots,  x_m, F, \mb{S}) = 
\forall \mb{x} ~ \hat{\eta} (\mb{x}, F, \mb{S}),
\label{polyMaxPsi}
\end{equation}
where $\hat{\eta}$ is a conjunction of Horn clauses
($\hat{\eta}$ captures the feasibility of solution \bf{S}) 
and 
$\mb{x} = (x_1, \cdots, x_m)$. 
Note that $\hat{\eta}$ needs to capture two types of conditions 
(an example with such conditions is provided in the next section):
\benum
\item[(a)]
\it{Global} conditions (those that apply over all \bf{w} tuples): Such
conditions express the fact that the solution (\bf{S}, F) as a whole is a
feasible solution to \bf{A}. One such condition is $|F| \geq K$
mentioned above. And 
\item[(b)]
\it{Local} conditions: The ones that are specific
to a given \bf{w} --- if $F(\mb{w})$ is true, that is. 
\eenum
Thus $\hat{\eta}$ is a conjunction of these two types of conditions.
The global conditions\footnote{
Observe that $\hat{\eta}_1$ captures the cardinality condition $|F| \geq
K$.  ~ To represent this, we can define a first-order relation $G$ of
arity $k$ over
the universe A of \bf{A} such that $|G| = K$ and $F(\mb{w})$ is true
whenever $G(\mb{w})$ is. ~Then we need to represent
the fact that $|F| \geq |G|$, which can be characterised as $\forall
\mb{w} ~ G(\mb{w}) \longrightarrow F(\mb{w})$. }
can be written as $\hat{\eta}_1$, and the local conditions as 
$\forall \mb{w} ~ F(\mb{w}) \longrightarrow \hat{\eta}_2 (\mb{x}, \mb{w},
\mb{S})$. 
~So $Q$, the decision problem, can be written as
$\exists \mb{S} ~ \exists F ~ \forall \mb{x} ~ \hat{\eta}_1 \wedge 
\left[ \forall \mb{w} ~ 
F(\mb{w}) \longrightarrow \hat{\eta}_2(\mb{x}, \mb{w}, \mb{S}) \right]$. 
In prenex normal form, 
\begin{eqnarray}
Q  & \equiv &
\exists \mb{S} ~ \exists F ~ \forall \mb{w} ~ \forall \mb{x}
~ \hat{\eta}_1 \wedge \left[ F(\mb{w}) \longrightarrow 
\hat{\eta}_2(\mb{x}, \mb{w}, \mb{S}) \right] \nonumber \\ [5pt]
& \equiv &  
\exists \mb{S} ~ \exists F ~ \forall \mb{w} ~ \forall \mb{x}
~ \hat{\eta}_1 \wedge 
 \left[ \neg F(\mb{w}) \vee \hat{\eta}_2(\mb{x}, \mb{w}, \mb{S}) \right]
\label{polyMaxDecisionEtaCap}
\end{eqnarray}

If $\hat{\eta}_1$ and $\hat{\eta}_2$ are each 
 a conjunction of Horn clauses, then so is the formula in 
(\ref{polyMaxDecisionEtaCap}). 
If we let $\eta (\mb{x}, \mb{w}, \mb{S}, F) = 
 \hat{\eta}_1 \wedge 
 \left[ \neg F(\mb{w}) \vee \hat{\eta}_2(\mb{x}, \mb{w}, \mb{S}) \right]
$, then 
(\ref{polyMaxDecisionEtaCap}) can be rewritten in 
 ESO Horn $\Pi_1$ form as
\begin{equation}
Q \equiv 
 \exists \mb{S} ~ \exists F ~ \forall \mb{w} ~ \forall \mb{x} ~
 \left[ \eta (\mb{x}, \mb{w}, \mb{S}, F) \right]. 
\label{polyMaxDecisionEta}
\end{equation}

To express the optimal solution value for $Q^{\prime}$, we maximise over all
feasible solutions $\mb{S}$ --- and for each solution, count the
number of $\mb{w}$ tuples for which the relation $F(\mb{w})$
and $\psi(\mb{w}, \mb{S})$ hold\footnote{
Out of the four possible cases 
(i) $(\forall \mb{x} \hat{\eta}) \wedge F(\mb{w})$, ~ 
(ii) $(\forall \mb{x} \hat{\eta}) \wedge \neg F(\mb{w})$, ~ 
(iii) $(\neg \forall \mb{x} \hat{\eta}) \wedge F(\mb{w})$, ~ and
(iv) $(\neg \forall \mb{x} \hat{\eta}) \wedge \neg F(\mb{w})$, 
cases (ii) and (iv) must be disregarded since $F$ is false.  Case (iii)
should also be disregarded since it violates the feasibility condition
$\forall \mb{x} \hat{\eta}$.}:
\begin{equation}
opt_{Q^{\prime}} (\mb{A}) = \max_{\mb{S},F} \vert \left\{ \mb{w}:
(\mb{A}, \mb{S},F) \models [ \forall \mb{x} ~ \hat{\eta}(\mb{x},
\mb{w},\mb{S}) ] \wedge F(\mb{w}) \right\} \vert.
\label{polyBdMaxOpt1} 
\end{equation}

In (\ref{polyBdMaxOpt1}),
$\forall \mb{x} ~ \hat{\eta}(\mb{x}, \mb{w},\mb{S})$ represents the
feasibility of the given instance $\mb{A}$.

If $\mb{S}$ and $F$ can be represented by a single sequence of
relations $\mb{T} = $ \\
 $(S_1, S_2, \cdots S_p, F)$, the optimal
solution value can be expressed as
\be
opt_{Q^{\prime}} (\mb{A}) = \max_{\mb{T}} \vert \left\{ \mb{w}:
(\mb{A}, \mb{T}) \models \forall \mb{x} ~ \eta
(\mb{x},\mb{w},\mb{T}) \right\} \vert
\label{optMax} 
\ee
where $\eta (\mb{x}, \mb{w},\mb{T}) = \hat{\eta} (\mb{x},
\mb{w},\mb{S}) \wedge F(\mb{w})$.
~ ($\hat{\eta}$ and $\eta$ are quantifier-free.) 
Since $\hat{\eta}$ is Horn, so is $\eta$. 

Hence the proof. 
\end{proof}

\subsubsection{Example: Polynomially Bound Maximum Flow (Unit Capacities)}
\label{sec:polyBoundMaxEx}

In this section, we will see how the MAXFLOW problem with unit
capacities can be expressed in ESO, in $\Pi_1$ form.  
Given a source $s$ and a sink $t$, and a network $G$ containing directed
edges, we want to find the maximum flow that can be sent through the
network from $s$ to $t$.
Essentially we seek the maximum number of edge-disjoint paths from $s$
to $t$. 
Call this (polynomially bound) problem MAXFLOW$_{PB}$. 

We want to determine the maximum number of vertices $w$ (one dimensional
tuples) to which there is a flow from $s$, along the edge $(s,w)$ (if
such an edge exists) --- this will give us the value of the maximum flow
from $s$ to $t$.  Every $s-t$ edge-disjoint path can be considered as a
partial order on the set of vertices. 
We will use ideas similar to those used in the expression
for REACHABILITY \cite{papa}.

To represent the partial orders, introduce a second-order ternary
predicate $P(x,y,w)$ which holds iff $x \not= y$ and there is an
edge-disjoint path from $s$ to $w$ to $x$ to $y$, in the feasible
solution --- the path from $s$ to $w$, is of course, just a single edge.
(The ``main arguments" for $P$ are $x$ and $y$ --- $w$ is just an
additional reference.)
Thus we seek the maximum number of $w$'s such that $P(w,t,w)$ is true. 
The following expressions capture the properties of a feasible solution. 

(1) If $P(x_1,x_2,w)$ holds, then so does $G(s,w)$ --- that is, the edge
$(s,w)$ is defined in $G$: 
\begin{equation}
\begin{array}{rcl}
\phi_1  & \equiv &  \forall x_1 \forall x_2 \forall w 
    ~ P(x_1,x_2,w) \longrightarrow G(s,w)  \\ [2mm]
 & \equiv &  \forall x_1 \forall x_2 \forall w ~ \neg P(x_1,x_2,w) \vee G(s,w). 
\end{array}
\label{phi1}
\end{equation}

(2) An edge $(i,j)$ can be a part of only one $s-t$ disjoint path
(equivalently, only one $w-t$ edge disjoint path): 
\begin{equation}
\begin{array}{rcl}
\phi_2  & \equiv &  \forall i \forall j \forall w_1 \forall w_2 
~ P(i,j,w_1) \wedge P(i,j,w_2) \wedge G(i,j) \longrightarrow (w_1 = w_2)
\\ [2mm]
 & \equiv &  \forall i \forall j \forall w_1 \forall w_2 
~ \neg P(i,j,w_1) \vee \neg P(i,j,w_2) \vee \neg G(i,j) \vee (w_1 = w_2). 
\end{array}
\label{phi2}
\end{equation}

(3) $P$ is non-reflexive: 
\begin{equation}
\phi_3   \equiv  \forall y_1  \forall y_2 ~ \neg P(y_1,y_1,y_2). 
\label{nonReflex}
\end{equation}

(4) $P$ is transitive: 
\begin{equation}
\begin{array}{rcl}
\phi_4  & \equiv &  \forall u_1 \forall u_2  \forall u_3 \forall w_3 ~ 
 P(u_1,u_2,w_3) \wedge P(u_2,u_3,w_3) \longrightarrow P(u_1,u_3,w_3).
\\ [2mm]
 & \equiv &  \forall u_1 \forall u_2  \forall u_3 \forall w_3 ~ 
 \neg P(u_1,u_2,w_3) \wedge \neg P(u_2,u_3,w_3) \vee P(u_1,u_3,w_3). 
\end{array}
\label{transitive}
\end{equation}

(5) And finally, any two adjacent vertices in $P$ should also be
adjacent in $G$: 
\begin{equation}
\begin{array}{rcl}
\phi_5  & \equiv &   
\forall z_1 \forall z_2 \forall w_4 ~ P(z_1,z_2,w_4) \wedge ~ \forall z_3 ~
\neg  [P(z_1,z_3,w_4) \wedge P(z_3,z_2,w_4)] \longrightarrow G(z_1,z_2)
\\ [2mm]
 & \equiv  & 
\forall z_1 \forall z_2 \forall z_3 \forall w_4 ~ P(z_1,z_2, w_4) \wedge 
\neg  [P(z_1,z_3, w_4) \wedge P(z_3,z_2, w_4)] \longrightarrow G(z_1,z_2)
\\ [2mm]
 & \equiv  & 
\forall z_1 \forall z_2 \forall z_3 \forall w_4 ~ \neg P(z_1,z_2, w_4) \vee 
  [P(z_1,z_3, w_4) \wedge P(z_3,z_2, w_4)] \vee G(z_1,z_2)
\\ [2mm]
 & \equiv  &
\forall z_1 \forall z_2 \forall z_3 \forall w_4 
~ [\neg P(z_1,z_2, w_4) \vee P(z_1,z_3, w_4) \vee G(z_1,z_2)]
\\ [2mm]
&  & \wedge
~ [\neg P(z_1,z_2, w_4) \vee P(z_3,z_2, w_4) \vee G(z_1,z_2)]. 
\end{array}
\label{adjacency}
\end{equation}

Let $\displaystyle \Phi = \bigwedge_{i = 1}^5 \phi_i$. 

Observe that each $\phi_i$ ($1 \leq i \leq 5$) is a $\Pi_1$ Horn formula, as
required by Theorem \ref{thom:hornMax}. 
The optimal solution value to the given instance (network $G$,
represented by a structure \bf{A}), is given by 
\begin{equation}
opt_Q (\mb{A}) = \max_{P} \left\vert \left\{ w :
(\mb{A}, {P}) \models  ~ 
P(w,t,w) \wedge \Phi \right\} \right\vert. 
\end{equation}

\bf{Discussion}. 
In most such expressions as above, there are two types of conditions to
be expressed: (1) \it{Global} conditions (those that apply over all
\bf{w} tuples), such as the expression $\Phi$ above and (2) \it{Local}
conditions (the ones that are specific to a given \bf{w}), such as
$P(w,t,w)$ above.  The first (second) set of conditions correspond to
\it{constraints} (\it{objective function}) in a classical mathematical
programming framework. 

It is clear that MAXFLOW$_{PB}$ can be expressed with neither 
$\Sigma_0$ nor $\Sigma_1$ formulae.
In particular, without universal quantifiers, none of the five
properties --- expressions (\ref{phi1}) to (\ref{adjacency}) --- can be
expressed independently of the size of the instance. 

Consider property $\phi_2$, for instance.  Using only existential
quantifiers, one should enumerate the property individually for each
edge.  However, this will make the length of $\phi_2$ dependent on the
number of edges in the graph.  Hence we can conclude that 
\begin{prop}
The property, that an edge belongs at most one edge-disjoint $s-t$ path
in a solution, (and hence the MAXFLOW$_{PB}$ problem) can be expressed
with a $\Pi_1$ formula, but not with a $\Sigma_1$ formula.
\label{piNOTsigma}
\end{prop}

\subsubsection{MAXFLOW$\mb{_{PB}}$ is Complete for Polynomially Bound
Maximisation}\label{sec:maxPcomplete}

We can show that the MAXFLOW$_{PB}$ problem is complete for the class of
polynomially bound maximisation problems by reducing 
an instance $I$ of a general problem $Q^{\prime}$ in this class to an
instance $\mc{I}$ of MAXFLOW$_{PB}$. ~If $I$ is represented by a
structure \bf{A}, the optimal solution value to $I$ is given by Theorem
\ref{thom:hornMax}: 
\begin{equation}
opt_{Q^{\prime}} (\mb{A}) = \max_\mb{S} \vert \{\mb{w}: (\mb{A},
\mb{S}) \models \Phi \}
\vert, ~~  
\Phi = \forall \mb{x} ~ \eta(\mb{w}, \mb{x}, \mb{S})
\label{maxDef2}
\end{equation}
if $Q^{\prime} \in \mb{Q^{\prime}}$,
where $\mb{x}$, $\mb{A}$, $\mb{S}$ and $\eta$ are defined in
Table \ref{tab:defines}.
(Recall that $\eta$ is a conjunction of Horn clauses and
quantifier-free, and \bf{S} is a sequence of second order predicate
symbols.)

Let the arity of \bf{w} (\bf{x}) be $k$ ($m$) respectively. 
The different possible \bf{w} (\bf{x}) tuples are 
$\mb{w_i}$, $1 \leq i \leq n^k$
($\mb{x_j}$, $1 \leq j \leq n^m$), 
where $n$ is the cardinality of the universe $A$ of \bf{A}. 
~For a given $\mb{w_i}$, the expression for $\Phi$  in (\ref{maxDef2})
can be rewritten as 
\begin{equation}
\Phi (\mb{w_i}) = 
 \forall \mb{x} ~ \eta(\mb{w_i}, \mb{x}, \mb{S}) = 
 \bigwedge_{j=1}^{n^m}  \eta(\mb{w_i}, \mb{x_j}, \mb{S}). 
\end{equation}

Instance $\mc{I}$ consists of $n^k + 2$ vertices --- one for each
$\mb{w_i}$
tuple, as well as two additional vertices $s$ and $t$.
Add a directed edge with unit capacity from $s$ to each $\mb{w_i}$ vertex.
Add a directed edge with unit capacity from each $\mb{w_i}$ vertex to
$t$ iff $\Phi (\mb{w_i})$ holds. 

The reduction is polynomial time --- $O(n^k)$ time to create the
vertices, and $O(n^{k+m})$ time to add the edges. 
It is clear that instance $\mc{I}$ of MAXFLOW$_{PB}$ has a maximum flow
of $\alpha$ units from source $s$ to sink $t$ iff the optimal solution
value to $I$ in (\ref{maxDef2}) is also $\alpha$.

\subsubsection{A Problem in MAX$\mb{_P \Sigma_0}$ ?}

Within the class $\mb{Q^{\prime}}$ (see Table \ref{tab:defines}), 
let us define the class MAX$_P \Sigma_0$ (MAX$_P \Pi_1$)
as the class of polynomially bound maximisation problems that can be
expressed by a $\Sigma_0$ ($\Pi_1$) formula. 

Kolaitis and Thakur showed that MAX3SAT
is in MAX$_{NP} \Sigma_0$ (defined similar to MAX$_P \Sigma_0$), 
a subset of $\mb{N^{\prime}}$. 
~On the other hand, MAX2SAT is not known to be polynomially solvable
\cite{max2sat}, though the decision version, as to whether all clauses
are satisfiable, is well-known to be in \bf{P} \cite{gj}. 
~Results similar to MAX2SAT for both the maximisation and decision
versions are also known for HORNSAT (where every clause is required to
be a Horn clause) \cite{hornsat}. 

Towards the goal of obtaining a hierarchy within the polynomially bound
P-maximisation class, we need to exhibit a problem in MAX$_P \Sigma_0$. 
However, we have been unable to find such problem so far. 

We conjecture that as long as a successor relationship or a linear
ordering on the universe of a structure is necessary, a problem cannot
be expressed in MAX$_P \Sigma_0$ (since this will require a $\Pi_1$
expression).

\subsubsection{Hierarchy Within Maximisation}

We state the hierarchy within the polynomially bound maximisation class
without a formal proof (since it is clear from the argument below): If a
maximisation problem exists in the $\Sigma_0$ class (or the $\Pi_0$
class), then the $\Sigma_0$ class is strictly contained within the
$\Pi_1$ class. 

It is clear that the problem considered in Section
\ref{sec:polyBoundMaxEx}, MAXFLOW$_{PB}$, cannot be expressed with a
$\Sigma_0$ formula.  In particular, without quantifiers, none of the
five properties in Section \ref{sec:polyBoundMaxEx} --- expressions
(\ref{phi1}) to (\ref{adjacency}) --- can be expressed. 

From Section \ref{sec:maxPcomplete}, clearly MAXFLOW$_{PB}$ serves as a
complete problem for the MAX$_P \Pi_1$ class.
It would be desirable to obtain a complete problem for the MAX$_P
\Sigma_0$ class. 
An interesting observation is that the decision version of the weighted
MAXFLOW problem (where arc capacity can be any non-negative integer) is
complete for the class \bf{P} \cite{pComplete,immerman}. 

Another question to be answered is, is there a class MAX$_P \Sigma_1$
which is between MAX$_P \Pi_1$ and MAX$_P \Sigma_0$ ?

\subsection{Polynomially Bound P-Minimisation Problems} 
\label{sec:polyBoundMin}

For minimisation problems in $\mb{N^{\prime}}$, 
Kolaitis and Thakur \cite{KT94} proved the following
(see Table \ref{tab:defines} for a definition of $\mb{N^{\prime}}$ and 
Definition \ref{pie2sigma2} regarding $\Sigma_2$ formulae): 
\begin{thom}
A minimisation problem $Q^{\prime} \in \mb{N^{\prime}}$ if and only if there
exists a $\Sigma_2$ first order formula $\phi (\mb{w}, \mb{S})$  with
predicate symbols from the vocabulary $\sigma$ (of $\phi$) and the
sequence $\mb{S}$, such that for every instance $\mb{A}$ of $Q^{\prime}$,
\begin{equation}
opt_{Q^{\prime}} (\mb{A}) = \min_{\mb{S}} \vert \left\{ \mb{w}:
(\mb{A}, \mb{S}) \models \phi (\mb{w},\mb{S}) \right\} \vert. 
\end{equation}
\label{thom:KT94min}
\end{thom}
In other words, they showed that all polynomially bound NP-minimisation
problems fall in what can be called the MIN $\Sigma_2$ class.  In the
same paper, they also showed that this class is equivalent to the MIN
$\Pi_1$ class. 

A similar result can be shown for minimisation problems in
$\mb{Q^{\prime}}$ (the polynomial-time equivalent of $\mb{N^{\prime}}$):

\begin{thom}
Let $\mb{A}$ be a structure
(instance) defined over $\sigma$.  If $Q^{\prime}$ is a minimisation
problem in $\mb{Q^{\prime}}$, then the value of an optimal
solution to an instance $\mb{A}$ of 
$Q^{\prime}$ can be represented by
\begin{equation}
 opt_{Q^{\prime}} (\mb{A}) = 
 \min_{\mb{S},F} \vert \left\{ \mb{w}:
(\mb{A}, \mb{S}, F) \models  ~ 
 \forall \mb{x} ~ \tau 
\right\} \vert 
\label{minDef}
\end{equation}
where
$\tau = 
\eta(\mb{w}, \mb{x}, \mc{S}) \wedge
F(\mb{w})$, and
 $\mb{x}$, $\mb{A}$, $\mb{S}$, $\eta$ are defined as in
Table \ref{tab:defines}. 
(The symbol $F$ is a $k-$ary relation defined on the universe $|A|$ of
$\mb{A}$, since each $\mb{w}$ is $k-$dimensional.) 
\label{thom:hornMin}
\end{thom}

\begin{proof} 

The proof that $Q^{\prime}$ is polynomially bound is the same as in 
Theorem \ref{thom:hornMax}.

We start with Gr\"{a}del's Theorem (Theorem \ref{thom:decision}).  The
decision problem can be represented by 
an ESO Horn expression $\exists \mb{S} \psi$ where $\mb{S}$ is a
sequence of predicate symbols, and $\psi$ is a $\Pi_1$ first order
Horn expression where a successor relation is included in the vocabulary
of $\psi$. 

The analysis for the decision problem $Q$ is similar to the maximisation
case, except that one looks for at most $K$ tuples that satisfy the
feasibility condition $\eta (\mb{x}, \mb{W}, \mb{S})$\footnote{...... which
is the same as looking for at least $n^k - K + 1$ tuples that do not
satisfy $\eta (\mb{x}, \mb{W}, \mb{S})$.}. 

Thus for the minimisation version of the problem, an optimal value to an
instance \bf{A} can be written as
\be
opt_{Q^{\prime}} (\mb{A}) = \min_{\mb{S}} \vert \left\{ \mb{w}:
(\mb{A}, \mb{S}) \models  ~ \phi (\mb{w}, \mb{S})
\right\} \vert
\label{optMin1}
\ee
where $\phi (\mb{w}, \mb{S}) = 
        \forall \mb{x} ~ \eta (\mb{x},\mb{w},\mb{S})$. 

The $\mb{w}$ tuples 
can be considered as a $k-$ary relation $F$ such that $F(\mb{w})$ is
true if and only if $\mb{w} \in F$. ~Hence $\phi (\mb{w}, \mb{S})$ in
(\ref{optMin1}) should be modified to $\forall \mb{x} ~ \eta
(\mb{x},\mb{w},\mb{S}) \wedge F(\mb{w})$.  The number of tuples 
$\vert F \vert$ in $F$ should be minimised.

Again, out of the the four cases 
(i) $(\forall \mb{x} \eta) \wedge F(\mb{w})$, ~ 
(ii) $(\forall \mb{x} \eta) \wedge \neg F(\mb{w})$, ~ 
(iii) $(\neg \forall \mb{x} \eta) \wedge F(\mb{w})$, ~ and
(iv) $(\neg \forall \mb{x} \eta) \wedge \neg F(\mb{w})$, 
cases (ii) and (iv) should be disregarded since $F$ is false, and (iii)
violates the feasibility condition $\forall \mb{x} \eta$.
This leaves us with the following modification of (\ref{optMin1}): 
\begin{equation}
opt_{Q^{\prime}} (\mb{A}) = \min_{\mb{S},F} \vert \left\{ \mb{w}:
(\mb{A}, \mb{S}) \models  ~ 
(\forall \mb{x} \eta) \wedge F(\mb{w}) 
\right\} \vert
\label{optMin2}
\end{equation}
Note that 
$(\forall \mb{x} \eta) \wedge F(\mb{w}) =
\forall \mb{x} (F(\mb{w}) \wedge \eta)$. 
Since $\eta$ is Horn, so is 
$(F(\mb{w}) \wedge \eta)$. 

It may appear that the minimisation in (\ref{optMin2}) will always
result in an optimal value $|F|$ of zero, but since the minimum value is
taken only over all \it{feasible} solutions $(\mb{S}, F)$, the 
value obtained in (\ref{optMin2}) is correct.
 Hence the proof. 
(The minimisation over ``only feasible solutions $(\mb{S}, F)$" needs
further illustration and is provided below.)
\end{proof}

\bf{Illustration of 
Minimisation over ``only feasible solutions $(\mb{S}, F)$"}.
To illustrate this point, consider the SHORTEST PATH problem in 
Section \ref{sec:shortestPath}.  We attempt to minimise the number of
edges in a path from the source $s$ to the sink $t$.  If we minimise
over \bf{any} (\bf{S}, $F$) combination, obviously this minimum number
would be zero --- however, this would violate the feasibility condition
$\phi_1$ that there exists a path from $s$ to $t$.  Hence this ``zero"
solution is obviously infeasible. 

Consider another example, MIN SET COVER. ~We are given a \it{ground} set
$X$ and several subsets $Y_1$, $Y_2$, $\cdots$, $Y_q$ of $X$. 
~Let $C$ = $\{Y_1$, $Y_2$, $\cdots$, $Y_q\}$. 
~The problem is select a few (minimum number of) subsets $Y_i$ from $C$,
such that the union of the selected subsets is $X$.  We can associate a
unary tuple $w_i$ to each $Y_i$, such that the number of such $w$ tuples
in a solution is to be minimised. Let a second order predicate $S(w_i)$
determine if a certain subset $Y_i$ is chosen in a solution $S$.
~Obviously if we minimise over \it{all possible} $S$, the minimum number
of $Y_i$ subsets selected will be zero --- but then, such a solution is
clearly infeasible, since the union of the selected subsets (zero of
them!) is not equal to the ground set $X$.

\bf{Discussion}.  From Theorems \ref{thom:KT94max}-\ref{thom:KT94min} and 
\ref{thom:hornMax}-\ref{thom:hornMin}, the following can be observed 
in the case of polynomially bound optimisation problems:

While second-order expressions are able to distinguish clearly between
NP-maximisation and P-maximisation problems ($\Pi_2$ for the former and
Horn $\Pi_1$ for the latter), the distinction is less clear between
NP-minimisation and P-minimisation problems ($\Pi_1$ formulae in both
cases, the only distinction being the Horn clause requirement in the
P-minimisation case).

\subsubsection{Example: Shortest Path}\label{sec:shortestPath}

We now provide an example of a polynomially bound
P-minimisation problem, SHORTEST PATH$_{PB}$. 
~Assume that the edges have unit weight and are directed. 
The number of edges in the shortest path is to be minimised.
The decision version of this problem is easily represented as a
$\Sigma_1$ formula:
\begin{equation}
\exists x_1 \exists x_2 \cdots \exists x_k ~
G(s, x_1) \wedge G(x_1, x_2) \wedge \cdots \wedge G(x_{k-1}, x_k)
 \wedge G(x_k, t) 
\label{shPathDecision}
\end{equation}
where $s$ is the origin and $t$ is the destination. The above formula
says that there is a path from $s$ to $t$ of length $k+1$, and it is a
Horn formula.
($G(x,y)$ is true if there exists an arc $(x,y)$ in graph $G$.)
We have not used any second-order variables in (\ref{shPathDecision}),
hence the decision version is \it{FO (first order) expressible}.

\bf{Minimisation version}. 
A shortest path (or any path from origin to destination) represents a
partial order $P$ on the universe (the set of vertices) --- $P$ is
represented by a second order (SO)
binary predicate. Another SO 
binary predicate $S$ chooses which arcs in the network are in the
required path. Again, we will use ideas similar to those used 
for REACHABILITY \cite{papa}.
The following formulae express the properties of $P$ and $S$: 

 \renewcommand{\arraystretch}{1.15}
\begin{longtable}{rp{300pt}}
$\phi_1 \equiv$ & $P(s, t) \equiv \eta_1$  ~  (there exists a path from $s$ to $t$). \\

$\phi_2 \equiv$ & $\forall x ~ \forall y ~ \forall z ~ \eta_2, ~~
          \eta_2 \equiv [(P(x,y) \wedge P(y,z)) \rightarrow P(x,z)]$  
          ~ ($P$ is transitive.) \\

$\phi_3 \equiv$ & $\forall x ~ \forall y ~ \eta_3, ~~ 
    \eta_3 \equiv \neg P(x,x) \wedge 
         [(P(x,y) \rightarrow \neg P(y,x)]$ \\
         &  ($P$ is neither reflexive nor symmetric.) \\

$\phi_4 \equiv$ & $\forall x ~ \forall y ~ \eta_4, ~~ 
          \eta_4 \equiv S(x,y) \rightarrow [G(x,y) \wedge P(x,y)]$ 
  ~ (If an edge is chosen by $S$, then it has to be in the given
      graph $G$ and in the $s-t$ path $P$.) \\

$\phi_5 \equiv$ & 
      $\forall x ~ \forall y ~ \hat{\eta}_5$ with 
      $\hat{\eta}_5 \equiv P(x,y) \rightarrow 
      [S(x,y) \vee \exists z (P(x,z) \wedge S(z,y))]$, \\

   &  (Recursive definition of $P$ --- either there is an $(x,y)$ arc, or
      there exists a path from $x$ to $z$ and a $(z,y)$ arc.) \\

$\phi_6 \equiv$ & $\forall x ~ \forall y ~ \forall z ~ \eta_6$, ~~ 
         $\eta_6 \equiv [(S(x,y) \wedge S(z,y)) \rightarrow (x = z)]$  ~~
        (Predecessor is unique, hence there is a unique path $P$ from
          $s$ to $t$.) 
\end{longtable}

\medskip 

It can be shown that each $\eta_i$ ($1 \leq i \leq 6$) above is
equivalent to a Horn clause --- clauses with at most one positive literal
from the set of second order 
variables\footnote{A clause such as $P(x,z) \vee S(s,t)$ cannot be a Horn
clause, for instance.}  \{$P$,  $S$\} --- or a conjunction
of such clauses, as required by Theorems \ref{thom:decision} and 
\ref{thom:hornMin}.  The optimal solution value for instance
$\mb{G}$ can now be written in Horn $\Pi_1$ form as
\begin{equation}
opt (\mb{G}) = \min_{P,S} \left\vert \left\{ (p, q) :
(\mb{G}, {P}, S) \models  ~ 
\forall x ~ \forall y ~ \forall z ~ \bigwedge_{i=1}^6 \eta_i 
\wedge S(p,q) \right\} \right\vert. 
\end{equation}

\bf{Discussion}. 
Though we have not proved it, SHORTEST PATH$_{PB}$ could be one of those
problems where the decision version can be represented in $\Sigma_1$
form, but the optimisation problem is in $\Pi_1$ form.  It would be
interesting if this observation (hierarchy in terms of quantifier
complexity) could be proven or disproven. 

\section{Optimisation Problems in $\mb{P^{\prime}}$}\label{sec:nonPoly}

We next turn our attention to the class $\mb{P^{\prime}}$. ~This is the
set of all optimisation problems, \it{not necessarily polynomially
bound}, but the optimal solution can be computed within time
polynomial in the size of the input.   (These problems need not obey
Equation \ref{polyBoundDef}.) 

Zimand 1998 \cite{zimand98} generalised Theorems \ref{thom:KT94max} 
and \ref{thom:KT94min} to all
NP-Optimisation problems, not just those that are polynomially bound.
He showed that a $\Pi_2$ first-order formula captures the feasibility
conditions for any problem in this class, while the optimal solution
value can be represented by a maximisation (or minimisation) over
weighted tuples --- the tuples are similar to those used in
expressions (\ref{optMax}) and (\ref{optMin2}) for polynomially bound
problems, but now they are also assigned real number
weights.  The method of attaching weights to tuples has also been
discussed in Papadimitriou and Yannakakis 1991 \cite{papaYanna91}.

We will demonstrate (without a formal proof) that Zimand's result can be
extended to polynomial-time maximisation problems as well.  Zimand shows
that for any positive integer value $z$ for an optimal solution, we can
compute a set of weights $c_i$ that are
powers of two, such that $z = \sum_{i} c_i$.  However, for a
\it{given} optimisation problem $Q^{\prime}$, the weights on tuples are
\it{given quantities}, such as the arc capacities in a MAXFLOW problem
or the arc costs in a TRAVELLING SALESPERSON problem --- the weights are
part of the input. (Zimand makes no attempt to relate his computed
weights with the input weights.) 

Gr\"{a}del's Theorem states that any decision problem $Q \in \mb{P}$ can
be represented as 
\begin{equation}
Q \equiv \exists \mc{S} \psi. 
\label{nonPolyBnd} 
\end{equation}
A decision version of a maximisation problem asks if there is a
solution $\mc{S}$ to an instance $I$ (represented by a structure
$\mb{A}$) such that the objective function $f(\mb{A}, \mc{S}) \geq K$
where $K$ is a given constant.  

\bf{Motivation to attach weights to tuples}. 
If $I$ is a YES instance to $Q$, then $\psi$ must be
able to express the fact that $f(\mb{A}, \mc{S}) \geq K$ using a finite
structure, according to Gr\"{a}del's Theorem.  The quantity
 $f(\mb{A}, \mc{S})$,
though not polynomially bound in the size of $I$ any
more, is still a finite quantity.  For
a structure $\mb{A}$ with universe $A$, the number of $k-$ary
tuples possible is $|A|^k$, which is polynomial in the size of the
instance.  
In other words, $f(\mb{A}, \mc{S})$ need not be polynomially bound,
whereas the maximum number of $\mb{w}$ tuples should be 
--- this is in contrast to the problems in Sect. \ref{sec:polyBound}.
One way to capture a larger number ($f(\mb{A}, \mc{S})$) using a smaller
one (the number of $\mb{w}$ tuples) is by attaching weights to the
tuples. 

For example, in the MIN CUT problem (dual of MAX FLOW), the tuples
(arcs) are binary, and the weights of these tuples are the arc
capacities.  
In WEIGHTED MAX3SAT, the tuples (clauses) are ternary with a weight
attached to each clause.
In WEIGHTED MAXSAT, it is unknown how many literals are in each clause,
hence a unary tuple is commonly used \cite{KT94}.
~In TRAVELLING SALESPERSON (TSP), the weights on the binary tuples are the arc
costs. 
It may be undecided ahead of time how the optimal value to a
problem in $\mb{P^{\prime}}$ can be represented, as to which set of
tuples and their weights will be used ---
for example, the set of edges used in a solution to the TSP is unknown
until a solution is determined.
However, the number of such sets and their tuples are finite --- and the
weight of each tuple is a given quantity. 

Naturally, each set of tuples described above can be said to form a
relation $U_i$ over the universe of \bf{A}, and the set of all such
relations can be represented by \bf{U}.  ~ For $U_i \in \mb{U}$, its
weight $w(U_i)$ is defined as
\begin{equation}
w(U_i) = \sum_{\mb{w} \in U_i} ~ w(\mb{w}),
\label{weightSum} 
\end{equation}
where $w(\mb{w})$ is the given weight of tuple $\mb{w}$.  For example,
in a TSP instance with five vertices, each $U_i$ will contain five
tuples (the five arcs in the solution).  
However, in SHORTEST PATH, the number of tuples in $U_i$ depends on the
path (solution) used --- hence the cardinality of the different $U_i$'s
is not the same, since the number of arcs in each solution can vary. 

Furthermore, the universe $A$ should consist of values (such as vertex
indices in graphs) for the variables, as well as weights for the
tuples\footnote{This is a variant of Many-sorted Logic.}. 
A unary relation $C(x)$ --- sometimes known as a \it{hidden relation}
--- decides if a given variable is a \it{basic}
variable, or a weight for one of the tuples.  The universe $A$ of a
structure $\mb{A}$ will be of the form
\begin{equation}
A = \{a_1, a_2, \cdots a_n, w_1, w_2, \cdots w_m\}
\end{equation}
where the $a_i$'s are possible values for the basic variables and the
$w_j$'s are possible weights for tuples of basic variables. For any
variable $x_i$, the following expression $\phi_1 (x_i)$ should hold:
\begin{equation}
\phi_1 (x_i) \equiv 
[C(x_i)] \longleftrightarrow \bigvee_{j = 1}^{n} [x_i = a_j]. 
\label{basicOrTuple}
\end{equation}

Introduce a relation $R(x_0, x_1, x_2, \cdots x_k)$ which holds true iff
 $x_0$ is a weight for the tuple $\mb{w} = (x_1, x_2, \cdots x_k)$
--- hence $C(x_0)$ is false, and all other $C(x_i)$'s are true.  
In an instance, the variable $x_0$ is instantiated with a weight $w_i$
from $A$.

\subsection{Maximisation Problems}\label{sec:nonBoundMax} 

Reverting to Gr\"{a}del's expressibility in (\ref{nonPolyBnd}), 
since $\psi$ is in $\Pi_1$ ESO Horn form, 
it can be written as (just like the case for polynomially bound problems), 
\begin{equation}
\psi (\mb{x}, \mb{w_i}, \mc{S}) = \forall x_1 \forall x_2
\cdots \forall x_m ~ 
\hat{\eta}(x_1, x_2, \cdots x_m, \mb{w_i}, \mc{S}) = \forall
\mb{x} ~ \hat{\eta}(\mb{x}, \mb{w_i}, \mc{S})
\end{equation}
where $\mb{x} = (x_1, x_2, \cdots x_m)$ --- hence $\hat{\eta}$ should
express the fact that $w(U_i) \geq K$, and $\hat{\eta}$ should include
expressions for every $\phi_1 (x_i)$ in (\ref{basicOrTuple}). 
If a certain relation $U_i$ that satisfies the feasibility conditions
exists,  then
\begin{equation}
Q \equiv \exists\mc{S} ~ \exists U_i ~ \forall \mb{w} ~
\forall \mb{x} 
\left[ U_i(\mb{w}) \longleftrightarrow 
       \hat{\eta}(\mb{x}, \mb{w}, \mc{S}) \right],
\end{equation}
where $Q$ is the decision problem. 
From this, the value of the optimal solution (for the optimisation
problem $Q^{\prime}$) can be expressed as
 \begin{equation}
 opt_{Q^{\prime}} (\mb{A}) = \max_{\mc{S}, U_i} 
 \left\{ w({U_i}): 
 (\mb{A}, \mc{S}, U_i) \models 
 \forall \mb{w} \forall \mb{x} 
\left[ U_i(\mb{w}) \longleftrightarrow 
       \hat{\eta}(\mb{x}, \mb{w}, \mc{S}) \right]
 \right\}.
 \label{maxDefNonPoly}
 \end{equation}

$(\mb{A}, \mc{S}, U_i)$ above also satisifies expressions where
$U_i(\mb{w})$ and $\hat{\eta}(\mb{x}, \mb{w}, \mc{S})$ are false ---
however, since $U_i(\mb{w})$ is false, the weight of this tuple $\mb{w}$
will not be counted in $w(U_i)$. 

~Note that (\ref{maxDefNonPoly}) need not be a $\Pi_1$ Horn formula any
more, since the Horn property of $\neg \hat{\eta}(\mb{x}, \mb{w},
\mc{S})$ is unknown: 
\begin{equation}
\left[ U_i(\mb{w}) \longleftrightarrow 
       \hat{\eta}(\mb{x}, \mb{w}, \mc{S}) \right] 
\equiv
\left[U_i(\mb{w}) \vee \neg \hat{\eta}(\mb{x}, \mb{w}, \mc{S}) \right]
\wedge
\left[\neg U_i(\mb{w}) \vee \hat{\eta}(\mb{x}, \mb{w}, \mc{S}) \right].
\end{equation}

\subsection{Example: Weighted Matching}

Here, we provide an example of how WEIGHTED MATCHING
(optimisation version) can be expressed.  Given a graph \bf{G} with weights on
the edges, the objective is to \it{mark} certain edges such that the sum
of the weights on the marked edges is maximised, with the condition that
no two adjacent edges in \bf{G} can be marked. (In the context of this
problem, a \it{Matched edge} is a synonym for a \it{Marked edge}.)
An instance (structure) \bf{A}
consists of \\
(a) the universe $A$ (the union of the set of vertices and the set
of tuple-weights), \\
(b) a relation $G$ (the set of edges), \\
(c) a relation $C(x)$, which defines whether a variable is a vertex or
the weight of a tuple, \\
(d) and a ternary relation $R(x_0, x_1, x_2)$ that decides whether an
edge $(x_1, x_2)$ is assigned a weight of $x_0$. 

Let relation $U(v_i, v_j)$ be true if $(v_i, v_j)$ is a matched
edge.  Obviously it can be a matched edge only if the edge exists in the
given graph.  This is expressed by $\phi_0$ below. 
If edge $(v_i, v_j)$ is matched and $x$ is a vertex not in 
$\{v_i, v_j\}$, then an adjacent edge $G(x, v_i)$ (if it exists in the
given graph) cannot be matched.  This is expressed by $\phi_1$. The
three other expressions $\phi_2$, $\phi_3$ and $\phi_4$ perform the same
task.

$\phi_1 = U(v_i, v_j) \rightarrow G(v_i, v_j)$, 

$\tau = 
(x \not= v_i) \wedge (x \not= v_j) \wedge U(v_i, v_j)$,

$\phi_1 = 
\tau \wedge G(x, v_i) \rightarrow \neg U(x, v_i)$, 
~~ 
$\phi_2 = 
\tau \wedge G(v_i, x) \rightarrow \neg U(v_i, x)$, 

$\phi_3 = 
\tau \wedge G(x, v_j) \rightarrow \neg U(x, v_j)$, 
~~ 
$\phi_4 = 
\tau \wedge G(v_j, x) \rightarrow \neg U(v_j, x)$. 

\medskip

Let set of weights $B = 
\left\{ z \in A ~ \vert ~ \exists x \exists y ~ 
         U(x, y) \wedge R(z, x, y) \right\}$ --- however, since $B$ is a
set, if the same weight is assigned to two or more edges in $U$, only one of
them will be counted towards total edge weights.  Thus there is a need
to split $B$ into $B_i$ ($1 \leq i \leq m$, $m = $ number of edges in
the input) ---
a weight $w$ in $B_i$ occurs among $i$ edges in $U$.
~Hence\footnote{Issues such as quantifier complexity and Horn property
are irrelevant for the logic expressions in
(\ref{weightB1})-(\ref{edgeShortHand}).  Logic expressions are used here
for the sole purpose of defining $B_k$, $1 \leq k \leq m$.}
\begin{eqnarray}
B_1 & = &   
\left\{ z \in A ~ \vert ~ \exists x ~ \exists y ~ \forall u ~ \forall v ~ 
         \tau \wedge U(x, y) \wedge R(z, x, y) 
 \right\}, 
\label{weightB1} \\ [3pt]
\mbox{where } ~ \tau & = & 
\{ [(u \not= x) \vee (v \not= y)] \wedge U(u,v) \}
         \rightarrow \neg R(z, u, v).
\end{eqnarray}

(Note:  In the definition of $B_k$ below, $\exists_{i=1}^k x_i$ is a
shorthand for $\exists x_1 ~ \exists x_2 \cdots \exists x_k$.) 

In general, any $B_k$ ($1 \leq k \leq m$) can be expressed as
\begin{equation}
B_k  =    
\left\{ z \in A ~ \vert ~ \exists_{i=1}^k x_i ~ \exists_{i=1}^k y_i ~ \forall u ~ \forall v ~ 
   \bigwedge_{i=1}^k U(x_i, y_i) \bigwedge_{i=1}^k R(z, x_i, y_i) \wedge
\tau
 \right\},  
\end{equation}
where 
$ \tau = \tau_1 \wedge \tau_2, $ and, 
\begin{equation}
\tau_1  =  \left\{\bigwedge_{i=1}^k 
       \left[(u, v) \not= (x_i, y_i) \right]  \wedge U(u,v)
 \right\} 
 \rightarrow \neg R(z, u, v), 
\label{zIsAWeightForOnlyKedges} 
\end{equation}
\begin{equation}
\tau_2  =  \bigwedge_{i \not= j} (x_i, y_i) \not= (x_j, y_j) 
\label{kDiffEdges}
\end{equation}
\begin{equation}
\left\{ (x_i, y_i) \not= (x_j, y_j) \right\} \equiv 
\left\{  (x_i \not= x_j) \vee (y_i \not= y_j) \right\}. 
\label{edgeShortHand} 
\end{equation}
Expression (\ref{kDiffEdges}) says that there are $k$ distinct edges
$(x_i, y_i)$. Expression (\ref{edgeShortHand}) is an explanation of the
shorthand notation used in 
(\ref{zIsAWeightForOnlyKedges}) and 
(\ref{kDiffEdges}) --- that if two edges are different, then at least
one of their endpoints should be different. 

~The weight of relation $U$, $w(U)$, is computed as: 
\begin{equation}
w(U) = 
\left( \sum_{z \in B_1} z \right) + 
\left( 2 \sum_{z \in B_2} z \right) +  \cdots +
\left( m \sum_{z \in B_m} z \right). 
\end{equation}

Finally, $\Phi$ is the expression that a solution $U$ should satisfy,
and the optimal solution value is obtained by maximising over all such
solutions: 
\begin{eqnarray}
\Phi & = & \forall v_i ~ \forall v_j ~ \forall x ~ 
[C(v_i) \wedge C(v_j) \wedge C(x)] \rightarrow \bigwedge_{k=0}^4 \phi_k,
\\ [2pt]
 opt_{Q^{\prime}} (\mb{A}) & = & \max_{U} 
 \left\{ w({U}): 
 (\mb{A}, U) \models \Phi  \right\}. 
\end{eqnarray}

\section{Future Research}

The open question 
--- mentioned in the proof to Theorem \ref{thom:hornMax} --- 
of how to express
decision versions of optimisation problems in the $\Pi_1$ form specified
by Gr\"{a}del for problems in \bf{P} in Sect. \ref{sec:polyBoundMax} 
needs resolution. 
A formal proof is needed for the arguments in Sect.
\ref{sec:nonBoundMax}. Furthermore, Sect.  \ref{sec:nonPoly} studies
only maximisation problems --- research should be carried out for
minimisation problems as well.  Complete problems should be discovered
for the respective subclasses. 

Since the decision version of the weighted MAXFLOW problem (where arc
capacity can be any non-negative integer) is complete for the class
\bf{P} \cite{pComplete,immerman}, the optimisation version of weighted
MAXFLOW is likely to be a complete problem for $\mb{P^{\prime}}$ ---
this is yet to be proven.

\subsection*{Acknowledgements} 
We benefited from discussions with Dov Gabbay of Kings College (London),
the theoretical computer science group at the University of Leicester
(UK),
as well as with J. Radhakrishnan and A. Panconesi at TIFR (Mumbai). 
A preliminary version of the paper was presented at the Algorithms and
Complexity in Durham (ACiD 2005) workshop at Durham, UK.


\begin{thebibliography}{KMSV98}

\bibitem[{E. }91]{gradel91}
{E. Gr\"{a}del}.
\newblock {The expressive power of second order Horn logic}.
\newblock In {\em {STACS 1991: Proceedings of the 8th annual symposium on
  Theoretical aspects of computer science --- Lecture Notes in Computer Science
  280}}, pages 466--477. Springer-Verlag, 1991.

\bibitem[Fag74]{fagin}
R.~Fagin.
\newblock {Generalized first-order spectra and polynomial-time recognizable
  sets}.
\newblock In R.~Karp, editor, {\em {Complexity of Computations}}, pages 43--73.
  SIAM-AMS Proceedings (no.7), 1974.

\bibitem[GHR95]{pComplete}
R.~Greenlaw, H.~James Hoover, and W.L. Ruzzo.
\newblock {\em {Limits to Parallel Computation: P-Completeness Theory}}.
\newblock Oxford University Press, 1995.

\bibitem[GJ79]{gj}
M.R. Garey and D.S. Johnson.
\newblock {\em Computers and Intractability: A Guide to the Theory of
  NP-Completeness}.
\newblock Freeman (New York), 1979.

\bibitem[H\r97]{max2sat}
J.~H\r{a}stad.
\newblock {Some Optimal Inapproximability Results}.
\newblock In {\em {ACM-STOC 1997: Proceedings of the 29th ACM Symposium on the
  Theory of Computing}}, pages 1--10, 1997.

\bibitem[Imm99]{immerman}
Neil Immerman.
\newblock {\em {Descriptive Complexity}}.
\newblock Springer-Verlag, 1999.

\bibitem[KKM94]{hornsat}
R.~Kohli, R.~Krishnamurti, and P.~Mirchandani.
\newblock {The Minimum Satisfiability Problem}.
\newblock {\em SIAM Journal of Discrete Mathematics}, 7:275--283, 1994.

\bibitem[KMSV98]{Kmsv98}
S.~Khanna, R.~Motwani, M.~Sudan, and U.~Vazirani.
\newblock {On syntactic versus computational views of approximability}.
\newblock {\em SIAM Journal of Computing}, 28(1):164--191, 1998.

\bibitem[KT94]{KT94}
P.G. Kolaitis and M.N. Thakur.
\newblock {Logical Definability of NP-Optimisation Problems}.
\newblock {\em Information and Computation}, 115(2):321--353, December 1994.

\bibitem[KT95]{KT95}
P.G. Kolaitis and M.N. Thakur.
\newblock {Approximation Properties of NP-Minimisation Problems}.
\newblock {\em Journal of Computer and System Sciences}, 50:391--411, 1995.

\bibitem[Pap94]{papa}
C.H. Papadimitriou.
\newblock {\em {Computational Complexity}}.
\newblock Addison-Wesley (Reading, Massachusetts), 1994.

\bibitem[PR93]{pancoRanjan}
A.~Panconesi and D.~Ranjan.
\newblock Quantifiers and approximation.
\newblock {\em Theoretical Computer Science}, 107:145--163, 1993.

\bibitem[PY91]{papaYanna91}
C.H. Papadimitriou and M.~Yannakakis.
\newblock {Optimization, Approximation, and Complexity Classes}.
\newblock {\em Journal of Computer and System Sciences}, 43(3):425--440,
  December 1991.

\bibitem[Zim98]{zimand98}
M.~Zimand.
\newblock {Weighted NP-Optimisation Problems: Logical Definability and
  Approximation Properties}.
\newblock {\em SIAM Journal of Computing}, 28(1):36--56, 1998.

\end{thebibliography}
\end{document}